\documentclass[preprintnumbers,twocolumn,showpacs,amsmath,amssymb,floatfix,9pt,prd,
superscriptaddress,nofootinbib]{revtex4}
\usepackage{latexsym}
\usepackage{epsfig}
\usepackage{amssymb}
\begin{document}
\title{Exact solution   of a (2 + 1)
dimensional anisotropic star in Finch and Skea spacetime    }

\author{Piyali Bhar}
\email{piyalibhar90@gmail.com
 } \affiliation{
Department of Mathematics, Jadavpur University, Kolkata 700 032,
West Bengal, India}

\author{Farook Rahaman}
\email{rahaman@iucaa.ernet.in} \affiliation{Department of
Mathematics, Jadavpur University, Kolkata 700 032, West Bengal,
India}

\author{Ritabrata Biswas}
\email{biswas.ritabrata@gmail.com  } \affiliation{Department of
 Mathematics, BESUS, Howrah, India}

\author{Hafiza Ismat Fatima}
\email{hafizaismatfatima@yahoo.com} \affiliation{Department of
Mathematics, Govt. Degree college (W) Warburton, Nankana Sahib,
Pakistan}

\date{\today}

\begin{abstract}
\noindent   We provide a new class of interior solution of a (2 + 1)
dimensional anisotropic  star in Finch and Skea spacetime corresponding to the BTZ black hole.   We have developed the model by considering the MIT bag model EOS and a particular
ansatz for the metric function $ g_{rr}$ proposed by Finch and Skea \cite{FS}. Our model is free from central singularity
and satisfied all the physical requirements for the acceptability of the model.
\end{abstract}


\maketitle

 \section{Introduction}

Introduction of lower dimensional gravity, soon became an
interesting tool as many difficult  fundamental problems of $4~-$
dimensions became much much simpler in $(2+1)$ dimensions.  It did
enhance various understandings of gravity theories. Though
conventional realization of a black hole was first revealed in the
back ground of $4~-$ dimensional space time, a lower dimensional
analysis is preferred more to understand the different issues.
Banados, Teitelboim and Zanelli (hence BTZ) \cite{BTZ1}, in the
presence of negative cosmological constant, had constructed an
analytic solution representing the exterior gravitational field of a
black hole in $(2+1)$ dimensions which was opened up the possibility
of investigating many interesting features of black holes. Recently,
Rahaman {\em et al} \cite{Farook} have obtained    a different kind
of exact BTZ black hole solutions of Einstein's field equations in
$(2+1)$ dimensional anti-de Sitter back ground space-time in the
context of
    non-commutative geometry .
R. B. Mann and S. F. Ross had investigated under what circumstances
a disk of pressureless dust, actually a 3D analogue of
Oppenheimer-Snyder collapse, will collapse to the zero angular
momentum BTZ black hole \cite{Mann1}. The properties of such type of
collapse found to be the parallel process to the 4 dimensional ones.
Collapse to a point singularity occurs in finite proper time, and
the event horizon forms in infinite coordinate time, with an
infinite redshift. Martins et al \cite{Martins} considering the
collapse of dark energy fluids, had studied the collapse of an
anisotropic fluid with zero radial pressure. Irrespective of the
initial fluid nature they obtained a black hole as a result(in
$(2+1)$ dimensions). This was significantly different from higher
dimensional phantom energy collapse \cite{Rudra}. Cruz and
Zanelli\cite{Cruz} obtained an interior solution for an
incompressible fluid in $(2+1)$ dimensions and investigated the
bound on the maximum allowed mass of the resultant configuration. By
assuming a particular density profile, a class of interior solutions
corresponding to BTZ exterior was provided by Cruz et
al\cite{Cruz2}. Consideration of three-dimensional perfect fluid
stars in hydrostatic equilibrium(with polytropic EoS), matched to
the BTZ black hole exterior geometry was done by Paulo M. Sa.

The introduction of a negative cosmological constant for stellar
solutions with finite mass and finite radius, did match to an
exterior black hole geometry in a way similar to the situation in
four dimensions. The solution obtained by Sharma et al \cite{Sharma}
is regular at the center and it satisfies all the physical
requirements except at the boundary where the authors have proposed
a thin ring of matter content with negative energy density so as to
prevent collapsing. The discontinuity of the affine connections at
the boundary surface provide the above matter confined to the ring.
Such a stress-energy tensor is not ruled out from the consideration
Casimir effect for massless fields.

 Derivation of all perfect fluid solutions for
the static circularly symmetric space-time was done by Garcia \cite{Garcia}.
 The general solution is presented in the standard coordinate system   $\{t,r, \theta\}$ ,
and alternatively, in a system $-$ the canonical one $-$ with coordinates   $\{ t,N,\theta
\} $. The exact analytic form of the metric potentials, in this
formulation, can be obtained for any arbitrary choice of the density
profile or EOS of the matter content of the fluid source. In
particular, they presented a solution corresponding to a static
circularly symmetric perfect fluid source having constant energy
density, which (in the presence of a cosmological constant) might be
considered as analogous to the incompressible Schwarzschild interior
solution in $(3 + 1) $ dimensions. Since determination of the exact
analytic form of the solution in this formalism requires knowledge
about the EOS of the composition or the radial dependence of
energy-density, we find it worthwhile to adopt an alternative method
where the right hand side of the Einstein's field equations ($T_{ij
}$) will be governed by the geometry of the associated space-time
($G_{ij}$ ).

\section{Interior Space-time}
The metric for a static circularly symmetry Finch and Skea type star in $(2+1)$ dimensional spacetime can be written as
\begin{equation}
ds^{2}=-e^{2\nu(r)}dt^{2}+\left(1+\frac{r^{2}}{R^{2}}\right)dr^{2}+r^{2}d\theta^{2}
\end{equation}
Where $R$ is a curvature parameter and $t= constant$ hypersurface of the metric $(1)$ is parabolic in nature.\\

Let us assume that the energy momentum tensor  for the matter distribution at the interior of the star is of the form,
\begin{equation}
T_j^{i}=(\rho+p_r)u^{i}u_j-p_r g^{i}_j+(p_t-p_r)\eta^{i}\eta_j
\end{equation}
Where $u^{i}u_j~=~-\eta^{i}\eta_j=~1$ and $u^{i}\eta_i=0$.Here the vector $u^{i}$ is the fluid $3-$velocity and $\eta^{i}$ is the spacelike
vector which is orthogonal to $u^{i}$,$\rho$ is the energy density,$p_r$ and $p_t$ are respectively the radial and the transversal pressure of the fluid.
The Einstein field equation in $(2+1)$ dimension with cosmological constant $\Lambda(<0)$ for the spacetime given in equation $(1)$ together with
 the energy-momentum tensor given in equation $(2)$,assuming $G=c=1$ gives three independent equations as follows:
\begin{equation}
2\pi \rho+\Lambda=\frac{1}{R^{2}}\left(1+\frac{r^{2}}{R^{2}}\right)^{-2}
\end{equation}
\begin{equation}
2\pi p_r-\Lambda=\frac{\nu'}{r}\left(1+\frac{r^{2}}{R^{2}}\right)^{-1}
\end{equation}
\begin{equation}
2\pi p_t-\Lambda=\left(1+\frac{r^{2}}{R^{2}}\right)^{-1}\left(\nu'^{2}+\nu''-\frac{\nu'}{r^{2}+R^{2}}r\right)
\end{equation}
Let us consider the equation of state as
\begin{equation}
p_r=\frac{1}{3}(\rho-4B)
\end{equation}
where $B$ is the Bag constant.\\
Solving equation $(4)$ and $(6)$ with the help of equation $(3)$ we get,
\begin{equation}
\nu=\nu_0+\frac{1}{6}\ln\left(1+\frac{r^{2}}{R^{2}}\right)-(\Lambda+2\pi B)\frac{R^{2}}{3}\left(1+\frac{r^{2}}{R^{2}}\right)^{2}
\end{equation}
where $\nu_0$ is the constant of integration.\\
The space-time metric thus obtained is free from central singularity.
From equation $(3)$ we get the expression for density $\rho$ as,
\begin{equation}
\rho=\frac{1}{2\pi R^{2}}\left(1+\frac{r^{2}}{R^{2}}\right)^{-2}-\frac{\Lambda}{2\pi}
\end{equation}
Putting the value of $\nu$ in equation $(4)$ and $(5)$ the expression for radial pressure is obtained as,
\begin{equation}
p_r=\frac{1}{6\pi R^{2}}\left(1+\frac{r^{2}}{R^{2}}\right)^{-2}-\frac{1}{6\pi}(\Lambda+8\pi B)
\end{equation}
Using $(7)$ in $(5)$ we get the expression of $p_t$ as,
\begin{widetext}
\begin{equation}
p_t=\frac{\Lambda}{2\pi}+\frac{1}{2\pi}\left[\frac{1}{3R^{2}}\left(1+\frac{r^{2}}{R^{2}}\right)^{-3}\left(1-\frac{5r^{2}}{3R^{2}}\right)
+\frac{16}{9}(\Lambda+2\pi
B)^{2}r^{2}\left(1+\frac{r^{2}}{R^{2}}\right)-\frac{4}{3}(\Lambda+2\pi
B)\frac{\left(1+\frac{8r^{2}}{3R^{2}}\right)}{\left(1+\frac{r^{2}}{R^{2}}\right)}\right]
\end{equation}
\end{widetext}

\begin{figure}[htbp]
    \centering
        \includegraphics[scale=.3]{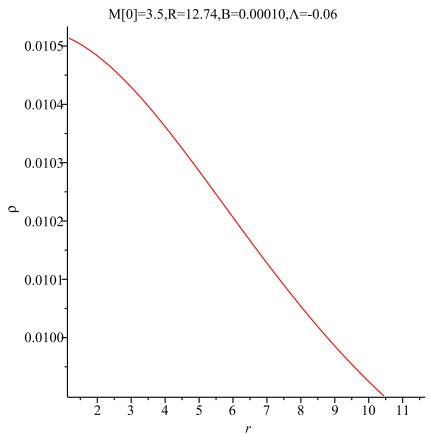}
       \caption{  Variation of the density of the stellar object in the
  interior region.}
    \label{fig:1}
\end{figure}

\begin{figure}[htbp]
    \centering
        \includegraphics[scale=.3]{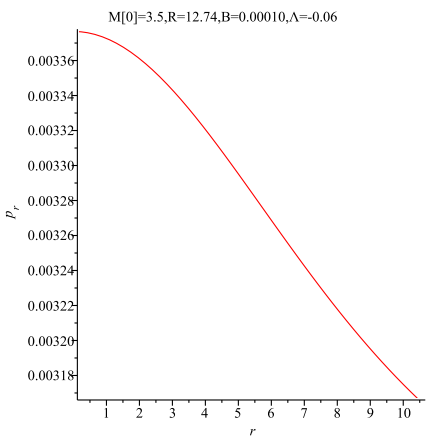}
       \caption{  Variation of the radial pressure of the stellar object in the
  interior region.}
    \label{fig:2}
\end{figure}

\begin{figure}[htbp]
    \centering
        \includegraphics[scale=.3]{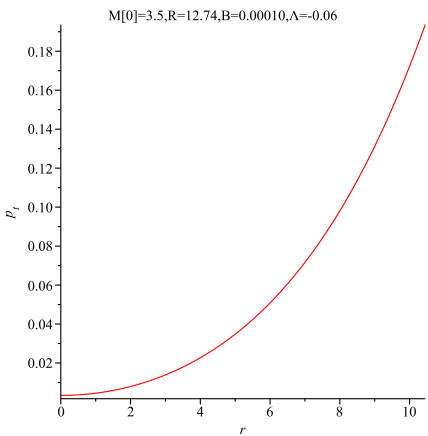}
       \caption{  Variation of the transversal pressure of the stellar object in the
  interior region.}
    \label{fig:3}
\end{figure}

\section{Exterior Space-time and Matching condition}

Let us match our interior solution to exterior BTZ Black hole metric,
\begin{equation}
ds^{2}=-(-M_0-\Lambda r^{2})dt^{2}+(-M_0-\Lambda r^{2})^{-1}dr^{2}+r^{2}d\theta^{2}
\end{equation}
Where the parameter $M_0$ is the conserved charged associated with asymptotic invariance under the time displacements.
Let us assume that $b$ is the radius of the star.Therefore,by the continuity of the metric function $g_{tt}$ and $g_{rr}$ at the boundary
of the star we get,

\begin{equation}
\left(1+\frac{b^{2}}{R^{2}}\right)^{1/3}e^{2\nu_0-\frac{2R^{2}}{3}(\Lambda+2\pi B)\left(1+\frac{b^{2}}{R^{2}}\right)^{2}}=(-M_0-\Lambda b^{2})
\end{equation}
and
\begin{equation}
\left(1+\frac{b^{2}}{R^{2}}\right)=(M_0-\Lambda b^{2})^{-1}
\end{equation}

Now at the boundary the pressure is zero ,i.e,$ p_r(b)=0$ which gives,
\begin{equation}
\frac{1}{6\pi R^{2}}\left(1+\frac{b^{2}}{R^{2}}\right)^{-2}-\frac{1}{6\pi}(\Lambda+8\pi B)=0
\end{equation}
Solving equation $(13)$ and$(14)$ we get,
\begin{equation}
R=\sqrt{\frac{b^{2}(-M_0-\Lambda b^{2})}{1+M_0+\Lambda b^{2}}}
\end{equation}
Which gives the value of the curvature parameter governing the value of the curvature spacetime of the metric given in equation $(1)$
and
\begin{equation}
B=\frac{1}{8\pi}\left[\frac{(1+M_0+\Lambda b^{2})(-M_0-\Lambda b^{2})}{b^{2}}-\Lambda\right]
\end{equation}
which gives the value of Bag Constant.

 Putting the values of $B$ and $R$ in equation $(12)$ one can easily
obtain the value of integration constant $\nu_0$.
Let us take the value of $\Lambda=-0.06$ and $M_0=3.5$ and radius of
the star $b=10.45$.Then from equation $(15)$ we get $R=12.74$ and from
equation $(16)$ we get $B=0.0001$.
 Note that while solving Einstein's equations as well as for plotting,   we have set c=G=1. However, if you keep these parameters
  from the beginning, the central density in our case turns out to be $ \rho_c = \frac{c^4  }{  G}\left[ \frac{1}{2\pi}(R^{-2}-\Lambda)\right]$.
  The Bag constant will be $ B_g=\frac{c^4  }{  G} \left[ \frac{1}{8\pi}\left\{\frac{(1+M_0+\Lambda b^{2})(-M_0-\Lambda b^{2})}{b^{2}}-\Lambda\right\}\right]$.
  Plugging G and c into relevant equations,  the
  value of the central density and  Bag constant $B$ turn out to be $ \rho_c = 1.5 ~\times ~ 10^{15}~ gm ~cm^{-3}$ and  $ B=83.7 ~MeV~ fm^{-3}$.

\section{Physical Condition}
For a physical meaningful solution we must have density and pressure are decreasing function of $r$. In our model,from equation $(8)$ and $(9)$,we have,
\begin{equation}
\frac{d\rho}{dr}=-\frac{2rR^{2}}{\pi(r^{2}+R^{2})^{3}}<0
\end{equation}
\begin{equation}
\frac{dp_r}{dr}=-\frac{2rR^{2}}{3\pi(r^{2}+R^{2})^{3}}<0
\end{equation}
At the point $r=0$ $$\frac{d\rho}{dr}=0~~~and~~~\frac{dp_r}{dr}=0$$\\
and $$\frac{d^{2}\rho}{dr^{2}}=-\frac{2}{\pi R^{4}}<0,\frac{d^{2}p_r}{dr^{2}}=-\frac{2}{3\pi R^{4}}<0$$
Which shows that density and pressure are decreasing function of $r$ and they have maximum value at the center and decrease radially
 outwards which have been shown in FIG.$1$ and FIG.$2$ respectively. \\

Now central density is obtained as,
\begin{equation}
\rho_0=\rho(r=0)=\frac{1}{2\pi}(R^{-2}-\Lambda)
\end{equation}
and radial and transverse pressures at center are obtained as,
\begin{equation}
p_r(r=0)=p_t(r=0)=\frac{1}{6\pi R^{2}}-\frac{1}{6\pi}(\Lambda+8\pi B)
\end{equation}

These  show that density and radial pressure both are regular at the center of the star. We have plotted density $(\rho)$ and radial
pressure $(p_r)$ in FIG.$1$ and FIG.$2$ respectively. The figures indicate  that density and pressure both are positive inside the stellar object.\\

 The equation of state parameters $\omega_r$ and $\omega_t$ are given
by the equations
\begin{equation}
\omega_r=\frac{p_r}{\rho}=\frac{\frac{1}{6\pi R^{2}}\left(1+\frac{r^{2}}{R^{2}}\right)^{-2}-\frac{1}{6\pi}(\Lambda+8\pi B)}
{\frac{1}{2\pi R^{2}}\left(1+\frac{r^{2}}{R^{2}}\right)^{-2}-\frac{\Lambda}{2\pi}},
\end{equation}
\begin{widetext}
\begin{equation}
\omega_t=\frac{p_t}{\rho}=\frac{\frac{\Lambda}{2\pi}+\frac{1}{2\pi}\left[\frac{1}{3R^{2}}\left(1+\frac{r^{2}}{R^{2}}\right)^{-3}
\left(1-\frac{5r^{2}}{3R^{2}}\right)
+\frac{16}{9}(\Lambda+2\pi
B)^{2}r^{2}\left(1+\frac{r^{2}}{R^{2}}\right)-\frac{4}{3}(\Lambda+2\pi
B)\frac{\left(1+\frac{8r^{2}}{3R^{2}}\right)}{\left(1+\frac{r^{2}}{R^{2}}\right)}\right]}{\frac{1}{2\pi
R^{2}}\left(1+\frac{r^{2}}{R^{2}}\right)^{-2}-\frac{\Lambda}{2\pi}}.
\end{equation}
\end{widetext}
We    plot these parameters  in $ FIG:4$ and $FIG:5$ respectively.
One can note that radial  EoS parameter lies within $0<\omega_r<1$,
however,  transverse  EoS parameter  increases with radial distance
gets the value which is  greater than unity.

The measure of anisotropy $ (\Delta=p_t-p_r)$  is given by,
\begin{widetext}
\begin{equation}
p_t-p_r=\frac{1}{2\pi}\left[-\frac{8r^{2}}{9R^{4}}\left(1+\frac{r^{2}}{R^{2}}\right)^{-3}+\frac{16}{9}(\Lambda+2\pi
B)^{2}r^{2}\left(1+\frac{r^{2}}{R^{2}}\right)+\frac{4}{3}(\Lambda+2\pi
B)\left\{1-\frac{\left(1+\frac{8r^{2}}{3R^{2}}\right)}{\left(1+\frac{r^{2}}{R^{2}}\right)}\right\}\right]
\end{equation}
\end{widetext}
 We have plotted the measure of anisotropy in FIG.$6$. Figure indicates  that the anisotropy is directed outwards.

\begin{figure}[htbp]
    \centering
        \includegraphics[scale=.3]{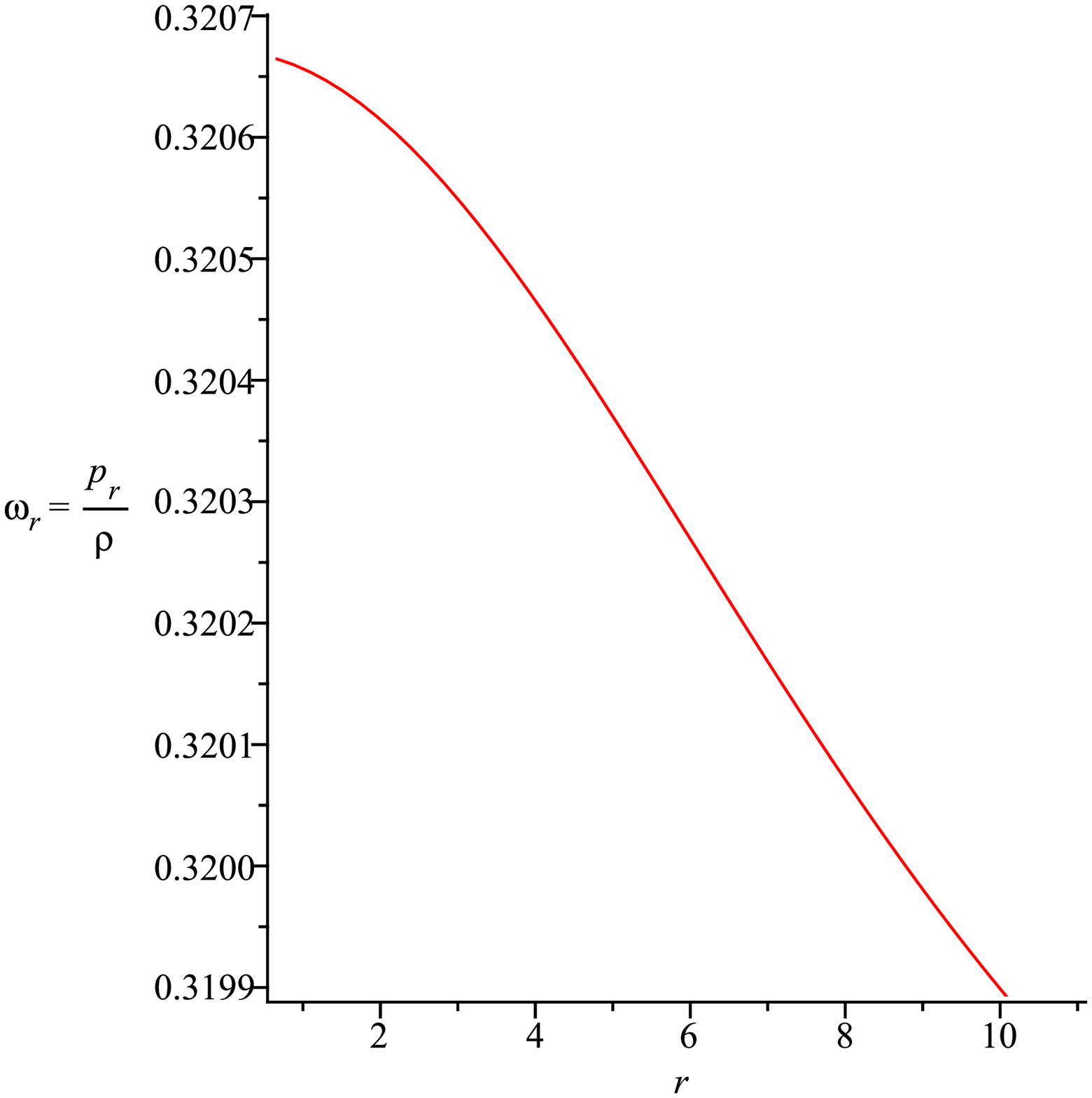}
       \caption{  Variation of equation state parameter $\omega_r$ against the radius $r$ is
            shown in the figure.}

    \label{fig:4}
\end{figure}
~\\
\\
\\
~

\begin{figure}[htbp]
    \centering
        \includegraphics[scale=.3]{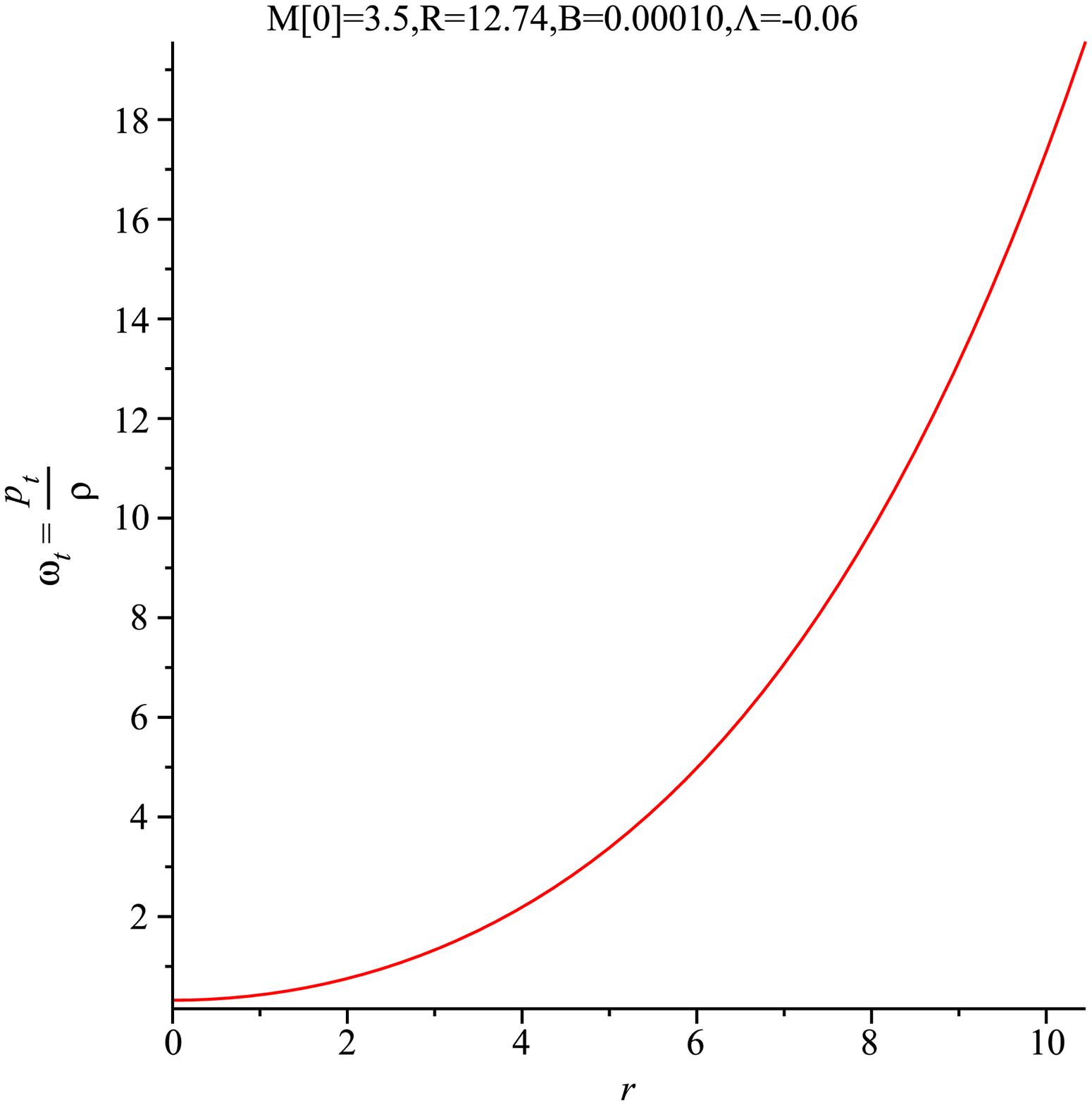}
       \caption{  Variation of equation state parameter $\omega_t $ against the radius $r$ is
            shown in the figure.}

    \label{fig:5}
\end{figure}

\begin{figure}[htbp]
    \centering
        \includegraphics[scale=.3]{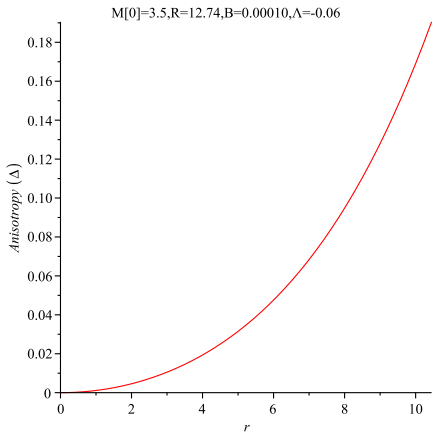}
       \caption{Anisotropy of the stellar object in the
  interior region.}
    \label{FIG:6}
\end{figure}

Now at the point $r=0,~~~\Delta=0$ which gives the physically meaningful solution.\\

We must have $-M_0-\Lambda b^{2}>0 $ for appropriate signature of the metric. This gives,
\begin{equation}
M_0 < -\Lambda b^{2}.
\end{equation}
Therefore,  we get an upper bound of $M_0$ from equation $(24)$.

\section{Energy Condition:}

It is well-known that that the null energy condition (NEC),weak energy condition(WEC) and the strong energy condition
(SEC) will be satisfied if and only if the following four equations holds simultaneously at every point within the source:
\begin{equation}
\rho +p_r \geq 0
\end{equation}
\begin{equation}
\rho \geq 0
\end{equation}
\begin{equation}
\rho +p_t \geq 0
\end{equation}
\begin{equation}
\rho+p_r+2p_t \geq 0
\end{equation}

Employing all the energy condition at the center $(r=0)$ We note that the first condition will be satisfied if
\[\frac{2}{3 \pi R^{2}}-\frac{2\Lambda}{3\pi}-\frac{4B}{3} \geq 0.\]
The second condition will be satisfied if
\[\frac{1}{R^{2}}\geq \Lambda~and~ \frac{1}{8\pi}\left(\frac{1}{R^{2}}-\Lambda\right)\geq B. \]
The third condition will be satisfied if
\[\frac{2}{3\pi}\left[\frac{1}{R^{2}}-(\Lambda+2\pi B)\right]\geq 0.\]
The fourth condition will be satisfied if
\[\frac{1}{\pi R^{2}}-\frac{\Lambda}{\pi}-4B \geq 0.\]
The assumed and estimated values of
$\Lambda, M_0, b, R, B$ are consistent with the  equations
$(25),(26),(27),(28)$ at the origin.
  We have plotted the L.H.S of
equations $(25),(26),(27)~and~(28)$ in $ FIG.7$ and from the figure
it is clear that all
 the energy conditions are satisfied.

\begin{figure}[htbp]
    \centering
        \includegraphics[scale=.3]{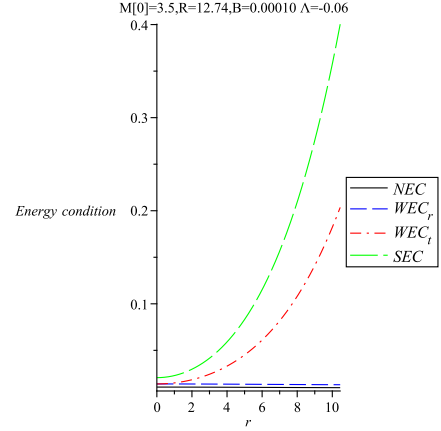}
       \caption{ the Variation of the energy of the stellar object in the
  interior region.}
    \label{fig:7}
\end{figure}

\section{Some Features:}

\subsection{Mass Radius Relation}
The mass function$ M(r)$ of the stellar object within the radial distance $r$ is given by,
$$ m(r)=\int_0^{r}2\pi \rho \widetilde{r}d\widetilde{r}
=\frac{r^{2}}{2}\left[\frac{1}{r^{2}+R^{2}}-\Lambda\right]$$

\begin{figure}[htbp]
    \centering
        \includegraphics[scale=.3]{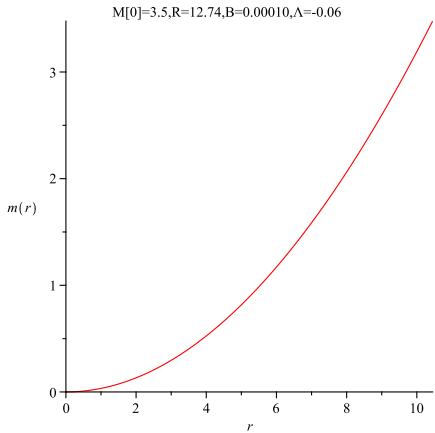}
       \caption{  Variation of mass of the stellar object in the
  interior region.}
    \label{fig:8}
\end{figure}

By using equation $(22)$ the upper limit of mass may be written as
\begin{equation}
m(r)_{max}\equiv M(b)=\frac{b^{2}}{2}\left[\frac{1}{b^{2}+R^{2}}-\Lambda\right]=\frac{1+M_0}{2}<\frac{1}{2}(1-\Lambda b^{2}),
\end{equation}
which gives,
$$\left(\frac{m}{r}\right)_{max}\equiv\frac{M}{b}<\frac{1}{2b}(1-\Lambda b^{2}).$$

\subsection{Compactness:}

The compactness of the stellar configuration is given by
\begin{equation}
u=\frac{m(r)}{r}=\frac{r}{2}\left[\frac{1}{r^{2}+R^{2}}-\Lambda\right].
\end{equation}
  To see the maximum allowance of the mass radius
ratio of our model we have plotted $\frac{m(r)}{r} vs r $ in
FIG.$9$. From the figure,  we notice that $\frac{m(r)}{r}$ is an
increasing function of 'r' and  the maximum value of
$\frac{m(r)}{r}$ is $0.3025~<\frac{4}{9}$. Thus $\frac{m(r)}{r}$ ratio lies within the standard limit of (3+1) dimensional case.  \\

\begin{figure}[htbp]
    \centering
        \includegraphics[scale=.3]{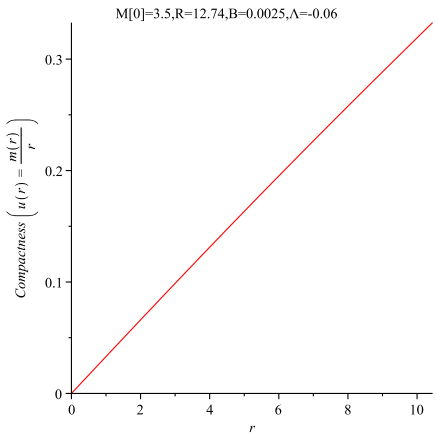}
       \caption{ compactness of the stellar object in the
  interior region.}
    \label{fig:9}
\end{figure}

\subsection{Surface Redshift:}

The surface red shift function $Z_s$ is given by
\begin{equation}
Z_s=[1-2u]^{-\frac{1}{2}}-1=\left[1-r\left(\frac{1}{r^{2}+R^{2}}-\Lambda\right)\right]^{-\frac{1}{2}}-1
\end{equation}
Which has been shown in figure $(10)$. The  maximum redshift  of our
(2+1) dimensional star of radius 10.45 turns out to be 0.729.

\begin{figure}[htbp]
    \centering
        \includegraphics[scale=.3]{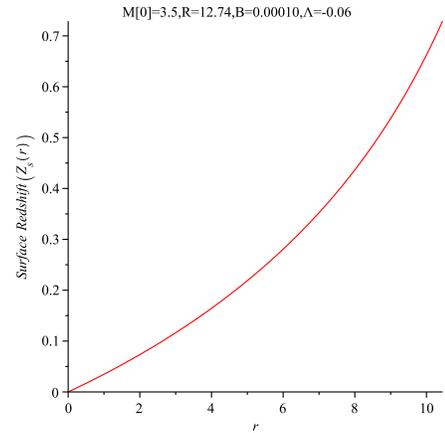}
       \caption{surface redshift function of the stellar object in the
  interior region.}
    \label{fig:10}
\end{figure}

\section{TOV Equation}

The Generalized Tolman-Oppenheimer-Volkoff (TOV) equation is of the
form \cite{leon}
\begin{equation}
-\frac{M_G(r)(\rho+p_r)}{r}e^{\frac{\nu-\mu}{2}}-\frac{dp_r}{dr}+\frac{2}{r}(p_t-p_r)=0
\end{equation}
where $M_G(r) $ is the gravitational mass within the radius $r$ and is given by,
\begin{equation}
M_G(r)=\frac{1}{2}re^{\frac{\mu-\nu}{2}\nu'}
\end{equation}
Substituting the value of $M_G(r)$ in equation $(25)$ we get,
\begin{equation}
-\frac{\nu'}{2}(\rho+p_r)-\frac{dp_r}{dr}+\frac{2}{r}(p_t-p_r)=0
\end{equation}

\begin{figure}[htbp]
    \centering
        \includegraphics[scale=.3]{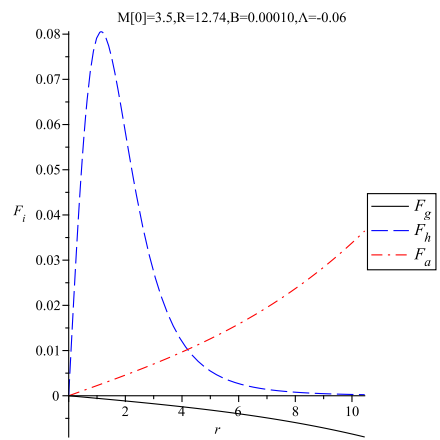}
       \caption{Three forces acting on the fluid are in static equilibrium.}
    \label{fig:11}
\end{figure}

The TOV equation in equation $(34)$ described the equilibrium of the stellar configuration under gravitational
force $F_g$,hydrostatic force $F_h$ and anisotropic stress $F_a$ within the stellar object so that
\begin{equation}
F_g+F_h+F_a=0
\end{equation}
where

\begin{widetext}
\begin{equation}
F_g=-\frac{\nu'}{2}(\rho+p_r)=\frac{2}{3\pi}\left[-\frac{r}{6R^{2}}\left(1+\frac{r^{2}}{R^{2}}\right)^{-1}+\frac{2}{3}(\Lambda+2\pi
B)r\left(1+\frac{r^{2}}{R^{2}}\right)\right]\left[\frac{1}{R^{2}}\left(1+\frac{r^{2}}{R^{2}}\right)^{-2}-(\Lambda+2\pi
B)\right]
\end{equation}
\end{widetext}
\begin{equation}
F_h=-\frac{dp_r}{dr}=\frac{2rR^{2}}{3\pi(r^{2}+R^{2})^{3}}
\end{equation}

\begin{widetext}
\begin{equation}
F_a=\frac{2}{r}(p_t-p_r)=\frac{1}{\pi
r}\left[-\frac{8r^{2}}{9R^{4}}\left(1+\frac{r^{2}}{R^{2}}\right)^{-3}+\frac{16}{9}(\Lambda+2\pi
B)^{2}r^{2}\left(1+\frac{r^{2}}{R^{2}}\right)+\frac{4}{3}(\Lambda+2\pi
B)\left\{1-\frac{\left(1+\frac{8r^{2}}{3R^{2}}\right)}{\left(1+\frac{r^{2}}{R^{2}}\right)}\right\}\right]
\end{equation}
\end{widetext}
In figure $FIG.11$  we have plotted $F_g,F_h,$ and $ F_a$ for a particular
stellar configuration. The figure indicates  that the
star is in static equilibrium under these three forces.

\section{Discussions}
We have presented a new class of solution in anisotropy corresponding to the BTZ exterior spacetime in $(2+1)$
 dimensional Finch and Skea spacetime . Our solutions are regular
  at center and   satisfy all the physical requirements .We have obtained the upper bound of
  the mass function. It was proved by Buchdahl that for maximally allowable mass-radius ratio $\frac{2M}{R}<\frac{8}{9}$ for (3+1) dimensional spacetime.
  For our model the maximally allowable mass radius ratio is
  $0.3025~<\frac{4}{9}$  which lies within the standard limit of (3+1) dimensional case.    Estimated value of Bag constant $B$ is 60-80 Mev $fm^{-3} $ for a $\beta$ equilibrium stable strange matter configuration \cite{MK,FR}.
  Plugging G and c into relevant equation the
  value of the Bag constant $B$ turns out to be B=83.7 Mev $fm^{-3} $ which is very close to the accepted
  value. However, one  concern of  this model is the transverse
  pressure. It  increases with radial distance as well as transverse
    EoS is greater than unity.  This  behavior is shown due
    to high compactness of the ultra-compact star where the density is
    above the nuclear density.

 \section{Acknowledgements}
FR  and RB would like to thank the Inter-University Centre for
Astronomy and Astrophysics (IUCAA), Pune, India, for research
facility.  FR is also grateful to UGC, Govt. of  India, for
financial support under its Research Award Scheme. PB is  thankful
to CSIR, Govt. of  India for providing JRF. RB thanks CSIR for
awarding Research Associate fellowship.

\end{document}